\documentclass[12pt]{article}

 \usepackage{amsfonts}
 \usepackage{amssymb}
 \usepackage{graphicx} \usepackage{epstopdf}
 \newcommand{\crlb}[1]{\label{#1}\\[2pt]} 
  
 \newcommand{\crld}[1]{\label{#1}}
 \newcommand{\eela}[1]{\quad\hbox{\scriptsize{#1}}\label{#1}\end{eqnarray}}
 \newcommand{\eelb}[1]{\label{#1}\end{eqnarray}}
 
 \newcommand{\newsecb}[2]{\section{#1}\label{#2}\setcounter{equation}{0}}

 \newcommand{\nolabels} {\def\eel{\eelb} \def\crl{\crlb} \def\newsecl{\newsecb}\def\bibiteml{\bibitem}\def\citel{\cite}\def\labell{\crld}}

\newcommand\publishversion{\nolabels\setlength{\textheight}{8.6in}\setlength{\oddsidemargin}{0in}
    \setlength{\textwidth}{6.3in}\setlength{\topmargin}{-0.3in}}

                 \def\fn{\footnote}
     \def\nm{\nonumber}   \def\be{\begin{eqnarray}}    \def\ee{\end{eqnarray}}
 \def\bi#1{\begin{itemize}\item[#1]}   \def\itm#1{\item[#1]}  \def\ei{\end{itemize}}  \def\eqn#1{(\ref{#1})}
 \def\tl#1{\tilde{#1}}  \def\^#1{\hat{#1}}

 \def\a{\alpha}      \def\b{\beta}   \def\g{\gamma}      \def\G{\Gamma}
 \def\d{\delta}      \def\D{\Delta}   
 \def\k{\kappa}      \def\l{\lambda}      \def\m{\mu}
 \def\f{\phi}        \def\F{\Phi}    \def\vv{\varphi}    \def\n{\nu}
 \def\j{\psi}            \def\r{\varrho}       
 \def\t{\tau}        \def\tht{\theta} \def\thh{\vartheta} 
               
 \def\w{\omega}      \def\W{\Omega}  

    \def\LL{{\mathcal L}}   
 \def\pa{\partial} \def\ra{\rightarrow}
  
 \def\dd{{\rm d}}  \def\bra{\langle}   \def\ket{\rangle}

 \def\iss{\ =\ }

 \def\fract#1#2{{\textstyle{#1\over#2}}}
 \def\ffract#1#2{\raise .2 em\hbox{$\scriptstyle#1$}\kern-.3em/
                 \kern-.2em\lower .15 em \hbox{$\scriptstyle#2$}}
 \def\fractje#1#2{{\scriptstyle{#1\over#2}}}
 \def\half{\fract12}  

 \def\ex#1{e^{\textstyle#1}}
  
\def\bmatrix{\begin{matrix}} \def\ematrix{\end{matrix}} \def\bpmatrix{\begin{pmatrix}}\def\epmatrix{\end{pmatrix}}
\def\bcenter{\begin{center}} \def\ecenter{\end{center}}

\def\lowerheightfig#1#2#3{\(\raise-#1\hbox{\includegraphics[height=#2]{#3}}\)}
\def\lowerwidthfig#1#2#3{\(\raise-#1\hbox{\includegraphics[width=#2]{#3}}\)}

\publishversion
\begin{document}
\bcenter 
{ \LARGE\textbf{ Black hole unitarity and \\[3pt]}}
{ \LARGE\textbf{ antipodal entanglement  \\[25pt] }}
{\large Gerard 't~Hooft}  \\[20pt]
Institute for Theoretical Physics \\[5pt]
\(\mathrm{EMME}\F\)   \\
Centre for Extreme Matter and Emergent Phenomena\\[5pt] 
Science Faculty\\ 
Utrecht University\\[5pt]
 POBox 80.195 \\
 3808TD, Utrecht  \\
The Netherlands  \\[10pt] 
e-mail:  g.thooft@uu.nl 
\\ internet:  http://www.staff.science.uu.nl/\~{}hooft101/ 
\ecenter
\vfil
 \noindent {\large\bf Abstract }

Hawking particles emitted by a black hole are usually found to have thermal spectra, if not exactly, then by a very good approximation. Here, we argue differently. It was discovered that spherical partial waves of in-going and out-going matter can be described by unitary evolution operators independently, which allows for studies of space-time properties that were not possible before. Unitarity dictates space-time, as seen by a distant observer, to be topologically non-trivial. Consequently, 
Hawking particles are only locally thermal, but globally not:  we explain why Hawking particles emerging from one hemisphere of a black hole must be 100 \% entangled with the Hawking particles emerging from the other hemisphere. This produces exclusively pure quantum states evolving in a unitary manner, and removes the interior region for the outside observer, while it still completely agrees locally with the laws of general relativity. 
Unitarity is a starting point; no other assumptions are made. Region $I$ and the diametrically opposite region $II$ of the Penrose diagram represent antipodal points in a PT or CPT relation, as was suggested before. On the horizon itself, antipodal points are identified.
A candidate instanton is proposed to describe the formation and evaporation of virtual black holes of the type described here. 

\vfil 
\eject
\setcounter{page}{2} 

\newsecl{Introduction}{intro}
	According to the classical picture of a black hole, it appears to be a sink that absorbs all matter aimed at it, without leaving a trace. The earliest descriptions of the quantum effects near a black hole, which lead to the marvellous conclusion that, due to vacuum polarisation, particles are emitted by a black hole\,\cite{SWH}, still suggested that quantum mechanics cannot prevent information to disappear. 
This, however, was quickly put in doubt\,\cite{GtHBH85}\cite{Susskind}. Precisely because the laws of thermodynamics apply to black holes\,\cite{BekHawk}\cite{SWH} , black holes must also be constrained to form quantum states with orthonormality and unitarity conditions. 

When this was realised, the author came with a possible scenario\,\cite{GtHgravshift}. Particles going in, do have an effect on the Hawking particles coming out, by modifying their quantum states, in spite of the fact that their thermodynamical distribution remains unaffected. The explanation of this effect is that particles going in interact with the particles going out. Ordinary, Standard Model interactions are too weak to determine the quantum states of the out-going particles in such a way that a unitary evolution operator could emerge, but the gravitational interactions, 
paradoxically, are so strong here that they dominate completely. Indeed, their effects can cause the relation between in- and out-going states to be unitary. Formally, the evolution operator, also called \(S\)-matrix, could be derived\,\cite{GtHSmatrix}, and the way it operates is quite reminiscent to the scattering operators provided by interactions in (super-) string theories. the black hole horizon then acts as the world sheet of a closed string.

We stressed that, in spite of the resemblance of this mathematical structure to string theory, this is not quite string theory, because the string slope parameter would be purely imaginary instead of real, and the string community paid only little attention. Yet, the author continues to defend the view that this should be seen as the most promising alley towards further understanding of nature's book keeping system. 
This is because the picture that emerges only works if matter is considered to be entirely \emph{geometrical} (just as in string theories), and thus, the condition that black holes should be entirely consistent with the laws of quantum mechanics, should be a powerful lead to guide us to a correct physical theory for all Planckian interactions.

Recently, it was found that the black hole back reaction can be calculated in a more systematic fashion. We did have a problem, which is that the gravitational force attributes \emph{too much} hair to a black hole. This is because the effects of the transverse gravitational fields could not yet be taken into account. Particles with too high angular momenta would contribute without bound, and this is obviously not correct. the search was for a method to give a transverse cut-off. A cut-off was proposed by Hawking et al\,\cite{HPS}, which is  a good starting point, while, being qualitative, it does not yet provide us with the exact expression for Hawking's entropy.

There is however an other question that can be answered: how can we separate the physical degrees of freedom near the horizon, so that each can be followed separately? This question was not posed until recently, and we found an astonishing answer\,\cite{GtHdiagonal}: we can diagonalise the information retrieval process completely, so that we can describe in closed form how information literally bounces against the horizon. Where, in previous calculations, a brick wall had been postulated\,\cite{GtHBH85}, a brick wall emerges naturally and inevitably when the diagonalised variables are used.

And there is more. We found that the degrees of freedom in region~$II$ of the Penrose diagram get mixed with the degrees of freedom in region~$I$. This is an inevitable element of the theory: gravitational deformations of space-time due to the gravitational fields of the particles going in and out, cause transitions from one region into the other. 

The question how this can be understood physically was not answered in our previous paper. Here, this question is answered.

\newsecl{Summary of the calculation of the effective bounce}{summary} \def\inn{{\mathrm{in}}}
\def\out{{\mathrm{out}}}

The details of this calculation were presented in Ref.~\cite{GtHdiagonal}. One begins by representing all possible forms of matter entering a black hole by their momentum distribution \(p^-_\inn(\tht,\vv)\). A single particle would give here a Dirac delta distribution. It is important that we must assume here that \emph{all} characteristics of in-going matter are duly registered by this distribution of the in-going momentum. Next, one considers the Hawking particles going out, by giving the out-going distribution \(p^+_\out(\tht',\vv')\). The canonically associated variables are the positions of the in- and out-going particles, \(u^+_\inn(\tht,\vv)\) and \(u^-_\out(\tht',\vv')\), where \(u^\pm\) are the light cone combinations of the Kruskal-Szekeres coordinates. A single particle wave function would be of the form\fn{Note that we use the metric convention (--,+,+,+), so that, here, \(p^+=\fract 1{\sqrt{2}}(k_r+E),\ p^-=\fract 1{\sqrt{2}}(k_r-E)\ ,\) while \(u^+=\fract 1{\sqrt{2}}(r+t),\ u^-=\fract 1{\sqrt{2}}(r-t)\). For particles going in, we have \(p^-<0,\ u^+>0\), while for the out-going particles, \(p^+>0,\ u^->0\). In Rindler \(I\), we have \(u^\pm>0\), in Rindler \(II,\ u^\pm<0\).}
	\be e^{i(p^-u^++p^+u^-)}\ . \eel{oneparticlewave}
If we have  distributions \(p^\pm(\tht,\vv)\), then the position operators \(u^\pm\) describe two single shells of matter, 
	\be u^+_\inn(\tht,\vv)\quad\hbox{and}\quad u^-_\out(\tht',\vv') \ , \eel{inoutshells}
obeying a simple commutator algebra with the momentum distributions.

In Refs.~\cite{GtHgravshift}\cite{GtHSmatrix}, the mechanism that relates `out' to `in' is worked out: gravitational interactions cause the out-particles to undergo a shift due to the momenta of the in-going ones, so that\fn{The details are also further explained in Ref.~\cite{GtHdiagonal}, where we show how the calculation is done in the Rindler limit. Here, we keep the finite size for the black hole, resulting in the extra term \(-1\) in Eq.~\eqn{greenfequ}.},
	\be u^-_\out(\W)=8\pi GR^2\int\dd^2\W f(\W,\W')p^-_\inn(\W')\ ,\qquad\W\equiv(\tht,\vv),\quad \W'=(\tht',\vv')\ , \eel{inoutgreen}
	where  \(R=2GM\) is the horizon radius. The Green function \(f\) obeys: 
	\be	\D_Sf(\W,\W')=-\d^2(\W,\W')\ ,\qquad \D_S=\D_\W-1=-\ell(\ell+1)-1\ ,\eel{greenfequ}
Units are chosen such that the Rindler limit, \(R\ra\infty,\ (\ell,m)/R\ra\tl k\) gives us ordinary flat space-time (\(\tl k\) is then the transverse component of the wave number).

	The new trick is that we expand both the momentum distributions \(p^\pm(\tht,\vv)\) and the position variables \(u^\pm(\tht,\vv)\) in terms of partial waves, \(Y_{\ell m}(\tht,\vv)\). In previous versions of this paper, the dependence on the horizon radius \(R\) was not worked out precisely. It turns out to be important to do this well. There is a difference in the \(u\) and the \(p\) variables in that the \(p\) variable is a \emph{distribution}, so it has dimension \(1/R^3\). We write (temporarily\fn{This algebra will be slightly modified in Eqs.~\eqn{antisigns} -- \eqn{elloddcomm}.})
	
		\be u^\pm(\tl x)\ra R\,u^\pm(\W)&,&\ p^\pm(\tl x)\ra {R^{-3}}\,p^\pm(\W)\ ; \labell{xtoOmega}\\
		\d^2(\tl x-\tl x')\ra R^{-2}\d^2(\W,\,\W')&,& [u^\pm(\W),\,p^\mp(\W')]=i\d^2(\W,\,\W')\ ;\labell{posmomalgebra}\\
		u^\pm(\W)=\sum\limits_{\ell,m}u_{\ell m}Y_{\ell m}(\W)&,& p^\pm(\W)=\sum\limits_{\ell,m}p^\pm_{\ell m}Y_{\ell m}(\W)\ . \\ \ 
		[u^\pm_{\ell m},\,p^\mp_{\ell' m'}]=i\d_{\ell \ell'}\d_{mm'}\ .&& \eel{ellmcomm}
	
We find that, at every \(\ell\) and \(m\), we have a complete set of quantum states that can be written in the basis \(|p^-_\inn\ket\) or \(|u^-_\out\ket\) or \(|u^+_\inn\ket\) or \(|p^+_\out\ket\), with each of these variables running from \(-\infty\) to \(\infty\). They obey the relations	
	\be u^-_\out=\frac{8\pi G/R^2} {\ell^2+\ell+1}p^-_\inn\ ;\qquad u^+_\inn=-\frac{8\pi G/R^2} {\ell^2+\ell+1}p^+_\out\ , \eel{upequs}
	Later, Sect.~\ref{oddell}, we shall see that we must limit ourselves to odd values of \(\ell\) only.

The wave functions \eqn{oneparticlewave} imply the Fourier relations (omitting the subscripts for short)
	\be \bra u^+|p^-\ket=\fract 1{\sqrt{2\pi}}\,e^{ip^-u^+}\ ,\qquad \bra u^-|p^+\ket=\fract 1{\sqrt{2\pi}}\,e^{ip^+u^-}\  . \eel{oneparticlewaves}
For every  \((\ell,\,m)\) mode,  we have these quantum states. For the time being, we now take \(\ell\) and \(m\) fixed.

Consider the time dependence, writing \(\t=t/4GM\). The variables \(p^-_\inn(t)\) and \(u^-_\out(t)\) increase in time as \(e^\t\), while \(u^+_\inn(t)\) and \(p^+_\out(t)\) decrease as \(e^{-\t}\). Because of this exponential behaviour, it is better to turn to familiar grounds by looking at the logarithms of \(u^\pm\) and \(p^\pm\). Then, however, their \emph{signs} \(\a=\pm\) and \(\b=\pm\) become separate variables. Write (for given values of \(\ell\) and \(m\)):
	\be u^+_{\ell,m}\equiv\a\,\ex{\r}\ ,\quad u^-_{\ell,m}\equiv\b\,\ex{\w}\ , \ee
We then have the time dependence
	\be \r(\t)=\r(0)-\t\ ,\qquad\w(\t)=\w(0)+\t\ . \ee
These ``shells" of matter bounce against the horizon, and the bounce is now generated by the wave equations \eqn{oneparticlewaves}. Note, however, that, in these equations, \(u^\pm\) and \(p^\pm\) will take both signs!

In Ref.~\cite{GtHdiagonal}, the wave functions are found to obey (in a slightly different notation)
	\be\j_\out(\b,\w)=\fract 1{\sqrt 2\pi}\sum_{\a=\pm}\int_{-\infty}^\infty\ex{\half(\r+\w)}\,\dd\r\,\ex{-\a\b\,i
			e^{\r+\w}}\j_\inn(\a,\r+\log\l)\ ,\nm\\
			\l=\frac{8\pi G/R^2} {\ell^2+\ell+1}\ .		\eel{psioutin}
Note that, since \(\hbar\) and \(c\) are put equal to one in this work, \(G\) is the Planck length squared, so that \(\l\) is dimensionless.
Next, the wave functions are expanded in plane waves in the tortoise coordinates \(\r\) and \(\w\):
	\be\j_{\inn}(\a,\r)=e^{-i\k\r}\,\j_\inn(\a)\ ,\qquad\j_\out(\b,\w)=e^{i\k\w}\,\j_\out(\b)\ , \ee
to find the Fourier transform of Eq.~\eqn{psioutin}:
	\be\j_\out(\b)=\sum_{\a=\pm}A(\a\b,\k)\,\j_\inn(\a)\ , \ee
with 
	\be A(\g,\k)=\fract 1{\sqrt{2\pi}}\G(\half-i\k)\,\ex{-\g\fract{i\pi}4-\g\k\fract{\pi}2}\ ,\eel{matrixcoeff} 
	where \(\g=\pm 1\).  Thus, we find that the waves scatter with scattering matrix\fn{Use was made of:
	\(\int_0^\infty\fract{\dd z}{\sqrt z}e^{\mp iz}z^{-i\k}=\G(\half-i\k)e^{\mp\fractje{i\pi}4\mp\fractje{\pi}2\k}\).}

	\be A=\pmatrix{A(+,\k)&A(-,\k)\cr A(-,\k)&A(+,\k)}\ , \eel{Amatrix}
and since 
	\be |\G(\half-i\k)|^2=\frac{\pi}{\cosh\pi\k}\ , \ee
we find this matrix to be unitary: 
	\be A\,A^\dag = \mathbb I\ . \ee
The diagonal elements, \(\g=+1\) show how waves interact when they stay in the same sector of the Penrose diagram. The off-diagonal elements switch from region \(I\) to \(II\) and back. Notice that the matrix elements keeping the particles in the \emph{same} sector are actually suppressed. Indeed, in the classical limit, \(\k\ra+\infty\),  we see that particles close to the horizon \emph{always} drag the out-going particles towards the other sector.

In Ref.~\cite{GtHdiagonal}, it is found that this scattering matrix gives the correct entropy of the horizon only if a cut-off is introduced in the angular partial waves: 
	\be\ell\le\ell_{\mathrm{max(M)}}\ . \ee
Concerning the present approach, Mersini\,\cite{Mersini} suggests that the cutoff in the angular momentum quantum number \(\ell\) can be argued using a ``quantum Zeno effect". However, her cut-off is a smooth one in the form of an exponent; for counting quantum states, this author considers a sharp cut-off more likely.

\newsecl{The domains in the Penrose diagram.}{penrosediag}
	Let us recapitulate what the findings reported about in the previous section mean physically. We presented in-going matter as a momentum distribution \(p_\inn^-(\tht,\vv)\) across the horizon, or equivalently, a cloud separated from the horizon by a distance function \(u^+_\inn(\tht,\vv)\). They obey the commutator algebra \eqn{posmomalgebra},
where \(u^+\) and \(p^-\) refer to the in-states, while \(u^-\) and \(p^+\) refer to the out-states.

\begin{figure}[h]\bcenter
	\includegraphics[width=300pt]{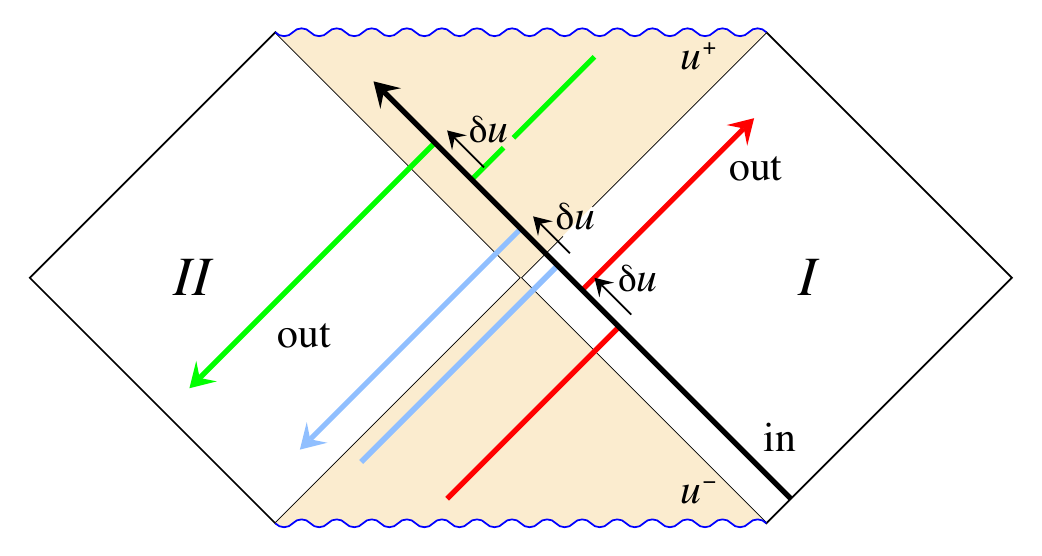}\ecenter \begin{quote}
	\begin{caption} { }Penrose diagram for Schwarzschild black hole, showing regions $I$ and $II$, a particle going in in region \# $I$, and particles going out in region $I$ and region $II$. The shift caused by the in-particle is the same in both cases, but in region $II$ the particle seems to go backwards in time. Since, in region $II$, the particle is shifted away from the event horizon, region $II$ experiences the same particle as a negative energy one, or an annihilated particle.\labell{penrosefig}
	\end{caption}\end{quote}
\end{figure}

	In performing the angular wave expansion of these functions, one clearly cannot avoid that both signs for the momentum and position amplitudes participate.  This is because in the original wave functions, \eqn{oneparticlewaves}, one cannot avoid that all signs for \(u^\pm\) and \(p^\pm\) contribute. One clearly must conclude that the scattering is only unitary if both signs for the position operator for the in-going and out-going shells of matter are included. The situation is illustrated in Fig~\ref{penrosefig}. Matter may enter in region $I$ or in region $II$ of the Penrose diagram. If the matter entering, or leaving, in region $II$ would have been omitted, the evolution would not have been unitary.

So what does the contribution from region $II$ mean? This question was not answered in Refs.~\cite{GtHdiagonal} or \cite{Mersini}. We note that unitarity is restored provided that Schwarzschild time \(t\) is used as the causal time parameter, so that, in region $II$, motion seems to go backward in the Penrose coordinates. Now, since we would like to refer to a single black hole here, we see only one clear option: region $II$ represents an other part of the same black hole. Therefore, we must identify a \(\mathbb Z_2\) mapping of the horizon onto itself, to represent both regions of the Penrose diagram.

There is one natural candidate: the antipodal mapping. This mapping has been proposed by Sanchez\,\cite{sanchezwhiting} in 1986.  One identifies
	\be (u^+,\,u^-,\,\tht,\,\vv)\qquad\hbox{with}\qquad(-u^+,\,-u^-,\ \pi-\tht,\,\pi+\vv)\ . \eel{antipodes}

For the description of classical particles, this identification has no observable consequences. We see in Fig.~\ref{penrosefig} that, a classical particle entering in region $I$ does not correspond to a classical particle, or antiparticle, or even the annihilation of an (anti-) particle in region $II$. Region $II$ is not reached at all, since also the signs of both \(u^+\) and \(u^-\) are switched. However, for quantum mechanics, this does make a difference. Wave functions have an extended support here, so a wave function in region \(I\) may have tails in region $II$. 

It is important to observe that the two points \eqn{antipodes} never come close together, so that no singularity is generated.

\newsecl{Entangled Hawking particles}{hawk}

Most significant is the effect this identification has for the Hawking particles. Sanchez and Whiting\,\cite{sanchezwhiting} state that Fock space cannot be used in their setting. We can however still use Fock space locally (although our description of in- and out-going matter to describe back reaction, indeed does not allow the use of Fock space). Assume now that we leave the black hole undisturbed for a while. When we then calculate the spectrum of Hawking particles, we see how these arise from vacuum fluctuations, and we can compute the distribution of particles as they would be seen by observers outside. In the author's reproduction of this calculation\,\cite{GtHSmatrix}, one finds that, if there is a vacuum as seen by the local observer at the origin of the \(u^\pm\) frame, this vacuum is an entangled state for the distant observer,
	\be |\j\ket=\sum_E e^{-\half\b_H E}|E,n\ket_I\,|E,n\ket_{II}\ . \eel{entangled}
Here, the states \(|E,n\ket_I\) describe the states with energy \(E\) and possible other quantum numbers \(n\) in region $I$, and we have the same in region $II$.
Eq.~\eqn{entangled} means that, \emph{if} an observer observes a particle with energy \(E\) and quantum numbers \(n\) emerging from a point \((\tht,\vv)\), (s)he will see \emph{exactly the same} particle, with the same energy \(E\) and quantum numbers \(n\) emerging at the antipodal point \((\pi-\tht,\,\pi+\vv)\). \emph{The Hawking particles emerging from one hemisphere of the black holes, are maximally entangled with the particles emerging from the other hemisphere.}

An interesting question is now: what is the Hawking entropy of such a black hole? Locally, all Hawking particles are indistinguishable from what they were in the standard black hole picture, Ref.\,\cite{SWH}, but now, one could argue that a rotation over an angle \(\pi\) in Euclidean space would bring us back to the original state, even though we travelled to the antipodes. This would suggest that we should multiply the temperature by 2 and divide the entropy by 2. We did have such a raise in temperature in a theory suggested by the author long ago\,\cite{GtHfactor2}, where region $II$ was identified with the bra states. We now claim however that any observer who would \emph{not} check the correlations between antipodal points, would only observe the standard Hawking temperature. Upon closer inspection, however, noticing the entanglement, the observer would conclude that temperature and entropy would be ill-defined concepts for a black hole. All its quantum states are now pure states.

The entanglement would be disrupted if we throw something into the black hole. As we see in the Penrose diagram, the pattern of Hawking particles would be shifted about, in opposite directions in the two regions.

We do emphasise that the new quantum states \eqn{entangled} of the Hawking particles do form a pure state, so, consequently, the information problem is completely resolved in the scenario suggested.

\newsecl{A gravitational instanton} {instanton}
The gravitational instanton that is strongly related to what was discussed above, is not the one studied most, which was elegantly derived by Eguchi and Hansen\,\cite{EguchiH}. Let us momentarily consider a Euclidean metric of the form
	\be \dd s^2=\dd r^2+a^2(r) \bigg(\dd\eta^2+(\sin\eta)^2\,\dd\thh^2+(\sin\eta)^2(\sin\thh)^2\,\dd\vv^2\bigg)\ , \eel{instantonmetric}
which has SO(4) rotational symmetry and is positive. The choice \(a(r)=r\) would give a 4 dimensional flat Euclidean space-time.

Now assume that, in stead of the usual condition \(r\ge 0\ ,\ \ a(0)=0\), we take 
	\be -\infty<r<\infty\ ,\qquad a(r)\approx \sqrt{r^2+\m^2}\ . \eel{boundary}
This space-time has two asymptotically flat regions, \(r\gg 0\) and \(r\ll 0\), unlike ordinary space-times. We correct for that by the antipodal identification of the points
	\be (r,\,\eta,\,\thh,\,\vv)\ \leftrightarrow\ (-r,\ \pi-\eta,\ \pi-\tht,\,\pi+\vv)\ . \eel{antipod}
Since the \(S_3\) sphere with the coordinates \((\eta,\,\thh,\,\vv)\) never gets a radius smaller than the parameter \(\m\), such an identification does not lead to any singularity anywhere. On the sphere \(r=0\), the antipodal points on the \(S_3\) sphere are identified. The metric can easiest be characterised by saying that, in ordinary Euclidean 4-space, we excise a sphere with radius \(\m\), and postulate that geodesics crossing that sphere continue outside the sphere at the antipodal point. Points inside the sphere are removed.

Naturally, one would ask which set of field equations allow for a solution with such a topology. In pure gravity one finds that the ansatz \eqn{instantonmetric} would necessarily send \(a(r)\) to zero somewhere, so that we have no solution resembling \eqn{boundary}. One might expect that matter fields could allow for solutions with this topology, but we can prove that ordinary scalar fields cannot do this job. Scalars with matter Lagrangian
	\be\LL_s(\f)=-\half\sqrt{g}(g^{\m\n}\pa_\m\,\f\pa_\n\f+V(\f))\ , \eel{scalar}
would lead to the equation
	\be 2a''+a(V(\f)+2{\f'}^2)=0\ , \ee
where \(\f'\) stands for \(\pa\f/\pa r\). Since the vacuum state is the state where \(V(\f)\) is lowest, and it must vanish at infinity, 
this enforces \(\pa^2 a/\pa r^2<0\), which would not allow the topology of \eqn{boundary}. In principle, adding a conformal term \(\l R\vv^2\) to the Lagrangian, or adding other types of matter fields, could produce such instantons, this we could not exclude.

The point we wish to make in this chapter is that the instanton \eqn{boundary} with the identification \eqn{antipod} would represent the formation and subsequent evaporation of a virtual black hole with antipodal identifications as in Eq.~\eqn{antipodes} in Section~\ref{penrosediag}. Since, at infinity, this instanton's metric approaches the flat metric faster than the Schwarzschild metric does, there is no external gravitational field noticeable at infinity. This, we interpret by saying that the total energy of the associated instanton tunnelling event vanishes.

\newsecl{On the condition that $\ell$ is odd}{oddell}
	When we say that region \(II\) is to be identified with the antipode of region \(I\), we say that the \(u\) variables as well as the \(p\) variables all obey
		\be u\,(\tht,\,\vv)=-u\,(\pi-\tht,\,\pi+\vv)\ .\eel{antipods}
Writing  \((\pi-\tht,\,\pi+\vv)\equiv -\W\,\) if \(\,(\tht,\vv)=\W\), the spherical harmonics \(Y_{\ell,m}\) obey
		\be Y_{\ell,m}(-\W)=(-1)^\ell \,Y_{\ell,m}(\W)\ ,\eel{antiell}
and therefore, only the odd values of \(\ell\) can contribute.

The algebra described by Eqs.~\eqn{posmomalgebra} -- \eqn{ellmcomm} is now replaced by:
	\be u^\pm(-\W)&=&-u^\pm(\W)\ ,\quad p^\pm(-\W)\iss -p^\pm(\W)\ ; \labell{antisigns}\\ \ 
	 [u^\pm(\W),\,p^\mp(\W')]&=& i\d^2(\W,\,\W')-i\d^2(\W,\,-\W')\ ,\labell{modcomm1}\\
	u^\pm(\W)&=&\fract 1{\sqrt 2}\sum\limits_{\ell=\mathrm{odd},\ -\ell\le m\le \ell}u^\pm_{\ell m}Y_{\ell m}(\W)\ ,\labell{oddpartialu}\\
	 p^\pm(\W)&=&\fract 1{\sqrt 2}\sum\limits_{\ell=\mathrm{odd},\ -\ell\le m\le \ell}p^\pm_{\ell m}Y_{\ell m}(\W)\ ,\labell{oddpartialp} \\	\ 	
	[u^\pm_{\ell m},\,p^\mp_{\ell' m'}]&=&	
	\bigg\{ \matrix {i\d_{\ell\ell'}\d_{m m'}& \hbox{if}& \ell\ =\ \hbox{odd} \cr
 0&\hbox{if}&\quad \ell\ =\ \hbox{even.}} \eel{elloddcomm}
(where we keep the \(\W\) integrals over the entire sphere).

What happens to a `spherical dust shell'?\,\cite{Kiefer} It would be a spherical wave with \(\ell=0\). The answer is that such a shell is not allowed in our formalism. The black hole horizon is covered with small domains (whose surface areas are of the order of the Planck size), with a sign function defined on each of these domains. The sign determines whether there is an object on \((\tht,\vv)\) or on \((\pi-\tht,\pi+\vv)\). It cannot be on both. A spherical dust shell would have positive \(u^+\) variables at all angles, which is not allowed. Instead, to mimic a dust shell, we should generate a density function, rather than a wave function, built from very many large (and odd) \(\ell\)-values. These would represent the particles, each on their tiny Planckian surface area, but, where one particle enters, an other particle cannot enter at its antipodal point. Only such dust shells, consisting of very many particles, would have a unitary evolution operator.
	
	Our algebra \eqn{xtoOmega} -- \eqn{upequs}, now modified into \eqn{antisigns} -- \eqn{elloddcomm}, is an inevitable consequence of the gravitational back reaction, and we showed that it connects regions \(I\) and \(II\) in a way that must be novel. In any case, it invalidates proposals to use both regions, \(I\) and \(II\) to describe the same physics\,\cite{Papadodimas}.  \(u^\pm\) and \(p^\pm\) each must have a single value according to the algebra, which can be positive or negative, and switch sign due to the gravitational back reaction, while the proposal to fold up the Penrose diagram so that region \(I\) would be equal to region \(II\) would force them to have two opposite signs at the same time. This would cause an unacceptable singularity at the centre of the Penrose diagram. Choosing these regions to describe points far separated in the \(\tht\) and \(\vv\) values avoids such a singularity. All this becomes manifest when doing the partial wave expansion as was shown in this paper.
	
	There may seem to be a problem. If the momentum variable \(p^-(\tht,\vv)\) would always be equal to \(-\,p^-\,(\pi-\tht,\,\pi+\vv)\), would this not mean that there is a complete pairwise cancellation among the momenta of the particles going in? And would this not be at odds with our freedom to choose the initial state without any antipodal symmetry? The answer to this is no. \(p^-\) does not at all represent the momenta of the in-going particles \emph{as seen by the outside observer}. First of all, it is weighed with a factor \(e^{-\t}\) with \(\t=t/4M\), so the momenta of late in-particles can be neutralised by exponentially tiny corrections for the momenta of particles going in much earlier, at the same spot. Secondly, a Bogolyubov transformation is required to project out the positive \(u\) values for region \(I\), and the negative \(u\) values for region \(II\). The wave function is a superposition of these two projections, that themselves need not to be correlated at all. So, in spite of the entanglement of the Hawking particles, the in-states at \((\tht,\vv)\) are in no way restricted by the in-states at \((\pi-\tht,\,\pi+\vv)\).

\newsecl{Discussion}{disc}

The central issue in our report is that the matrix \eqn{Amatrix} must be unitary, while it mixes the two regions of the Penrose diagram, so we have to conclude that region \(II\) also represents part of physical space-time. The only reasonable choice appears to be to identify region \(II\) with the antipodes.

One might want to search for arguments, such as stability arguments, to explain why \(\thh\approx 0\) would be favoured. This would go beyond what we planned to report here. We do note that, for Kerr and Kerr-Newman black holes, the axis of the extra rotation operator must coincide with the Kerr rotation axis, otherwise the mapping would lead to contradictions.

In our theory, something happens that has never been explicitly noticed: \emph{The arrow of time, in both regions, at all points near the horizon, must be taken to be the same as the external, Schwarzschild time, \(\t=t/4GM\).} This is contrary to standard practice\,\cite{HawkingEllis} and also contrary to the author's own earlier expectation. However, in order to keep unitarity, this is exactly what was done in the calculations reported about in this paper, and now we claim that calculations \emph{have} to be done this way. It implies that, in terms of the Penrose coordinates, the identification involves a \(PT\) transformation\fn{Presumably, this should be \(PCT\), but in our formalism the notion of antiparticles was not yet introduced; including electromagnetism in our formalism may well clarify this point.}: time in region \(II\) goes backwards. The gravitational shift effects that we incorporate, bring us from region \(I\) into region \(II\) and back, and we would not have unitarity if we used the time coordinate employed by local observers. Indeed, using the \(\t\) coordinate, we get the desired feature that our identification procedure commutes with time translations.

Note that, within our formalism, the `interior region' of a black hole disappears altogether, so that no problems with firewalls can arise.

Objections were raised by noting that a trapped region may emerge already in flat space-time, before the actual collapse takes place. So, the causal order of events will not be agreed upon by the different observers. This is true, but our priority goes to the construction of a unitary, causal evolution matrix for the black hole. If different observers, who cannot communicate anyway, disagree about causality, this will be an interesting discussion point but it will not invalidate our procedures. This does imply a change of mind w.r.t. our earlier theories and suggestions.

\newsecl{Conclusion}{conc}
An elegant way to phrase the new proposed theory, is to say that, when the first trapped region opens up, we can regard it as a very tiny black hole, coming into existence via a very tiny gravitational instanton. The fact that this tiny instanton has antipodal identifications is a minute modification of space-time structure inside the trapped region; then, when the region opens up wide, the new configuration grows together with it. A local observer near the horizon, sees both Penrose's regions \(I\) and \(II\), not realising that region \(II\) is a \((C)PT\) image of the antipodal part of the hole, since the same laws of physics apply there. This is why we say we do not violate general relativity with our identifications.

	We observed that antipodal identification of points on the horizon is inevitable if we want a unitary evolution operator. Formally, from the moment that a trapped space-time region forms, we must already identify antipodal points on the crossing point of future and past event horizons. Our point is that this remains invisible for `experiments', until one waits to see the quantum effects of a decaying and vanishing black hole. Even if our procedure seems to be quite natural, our familiar notions of space and time will have to be thoroughly revised. The advantage of our procedure of splitting things up in partial wave expansions, is that \emph{different partial waves are completely uncoupled}, so that we are left with very simple, finite-dimensional quantum mechanics for each wave, where one can exactly see what is going on.
	
The partial wave decomposition employed here should be distinguished from the usual partial wave decompositions in first- or second-quantized particle theories. We are forced to treat particles not as being point-like, but as forming a finite set of membranes that each take the shape of a partial wave. So, introducing a cut-off in \(\ell\) would not restrict total angular momenta of all particles any way. Although these partial waves have a classical appearance, we insist that they form legitimate representations of our operator algebra. They can be interpreted as a reformulation of the coordinates of all particles entering and leaving the black hole, a number that is roughly equal to \(R^2\) (in Planck units)\,\cite{Dvali}. The partial waves are then nothing but  a band-limited mode decomposition as was described in Ref~\cite{Kempf}.

Also, one should not expect a majority of Hawking particles to emerge with \(\ell\) values close to the Planck limit. To the contrary, as was emphasised by Dvali\,\cite{Dvali2}, Hawking radiation in practice is dominated by \(S\)-waves, with small tails in higher \(\ell\) modes, which are strongly suppressed by their Boltzmann factors. It is the micro-states that we arrange according to their \((\ell,\,m)\) values.

Our result can be characterised by saying that indeed black holes have hair, and using our procedures we can understand all its details:
\bi{--} Black holes have hair: one strand, on the average, for every Planckian surface elements of the horizon. Hair associated with the dynamical variable \(u^-_\out(\tht,\vv)\), sits on the expanding tortoise coordinate, so it grows exponentially. Hair associated to the variable \(u^+_\inn(\tht,\vv)\) shrinks exponentially in time. 
\itm{--} And then we have the sign variables \(\a(\tht,\vv)\), and \(\b(\tht,\vv)\). They have the features of a fermionic field on the scalp of the black hole, that is, they do not grow or shrink. As long as nothing falls into the black hole, the sign variable \(\a\) does not change with time, so it can be regarded as a conserved charge\,\cite{HPS}. The sign variable \(\b\), associated to the out-going particles, is also conserved, in principle, but it does not commute with \(\a\). For all practical purposes, \(\a\) and \(\b\) may be identified with the black hole micro-states, but remember that also the dynamical variables \(u^\pm\) contribute to the total entropy.
\ei

Let us emphasise, once again, that each partial wave decouples from all other partial waves, and this fact should be seen as a major discovery. It enables us to form a very simple picture of the structure of space-time at or near the Planck scale, without having to take our refuge in functional variables and integrals, which often obscure things. One finds that space and time have exciting features. The most important problem has always been that, at the black hole horizon, the local observer must allow for unlimited Lorentz boosts. These cause gravitational back reactions that are also unlimited. We now have a handle to cope with that situation: it was discovered that in-going particles exchange position operators with momentum operators, to turn into out-going particles.

Let us also emphasise that hardly any `approximation' has been made. Authors of other publications often belittle our results by claiming that it is merely a `classical approximation' or something like that. To the contrary, the algebra on which it is based is very compelling, having been derived from impeccable physical arguments. Indeed, the physics is very accurate as soon as we look at \(\ell\) values well below the maximal limit (at the Planck scale). How exactly to perform the cut-off at the maximal values of \(\ell\) is not precisely understood today, but the Planckian regime has not yet been well understood by anybody.

We believe that our results may lead to a superior view of the structure of space and time at the Planck scale. When a black hole is just about to be formed, a trapped region opens up, and as soon as that happens, the distant observer will be obliged to pairwise identify points on its boundary, in a \(PT\) invariant manner. Consequently, for the outside observer, the internal region disappears. Particles that enter the interior region, arranged in partial waves of energy and momentum, immediately re-emerge, with positions and momenta interchanged, as well as a switch in the arrow of time, while, most importantly, their quantum states remain pure, both for the inside observer as for the outside observer. The different partial waves do not mix.

There are many problems still wide open. One is the infinite black hole / Rindler limit. Where is the antipodal point in Rindler space? Does Rindler space have to be compactified just as the black hole horizon? On dimensional grounds, we do need a C-number \(R\) as in Eqs.~\eqn{xtoOmega}, \eqn{upequs} and \eqn{psioutin}.

Another problem is the fact that we were forced to treat the \(u\) variable here as the coordinate of a single `particle'. Although this is actually something more like a dust shell, one would still have expected that \(u^\pm\) should emerge as the coordinates of a second-quantised theory. This, emphatically, is not possible here (unitarity would get lost). What can be done in principle, is to treat the real number parameters \(u^\pm\) and \(p^\pm\) as sequences of binary digits instead. As the time parameter \(\t\) increases by an amount \(\log b\), where \(b\) is the base of the digital system used (for instance, \(b=2\)), the digits all move one step to the left or to the right, exactly as in a second quantised theory of fermions. We need a procedure of this sort in order to make contact with the Standard Model of the sub-atomic particles, but we would prefer a more elegant mathematical scheme.\vskip10pt

{\large \textbf{Acknowledgment}}\\[5pt]
	The author acknowledges useful discussions with A.~Ashtekar, P.~Betzios,  G.~Dvali, and B.~Whiting.

\end{document}